\documentclass[%
reprint,
superscriptaddress,
amsmath,amssymb,aps,
twocolumn,
citeautoscrip]{revtex4-1}

\usepackage{graphicx}
\usepackage{dcolumn}
\usepackage{bm}
\usepackage{color,soul}
\usepackage[dvipsnames]{xcolor}
\usepackage[colorlinks=true, linkcolor=blue, citecolor=blue,urlcolor=blue,breaklinks=true]{hyperref}
\usepackage{soul}
\usepackage{comment}
\usepackage{accents}
\usepackage{amssymb}
\newcommand{\ket}[1]{\left|{#1}\right\rangle}

\begin{document}
	
	\title{Unconditional violation of the shot noise limit in photonic quantum metrology}
	
	\author{Sergei Slussarenko}
	\affiliation{Centre for Quantum Dynamics, Griffith University, Brisbane, Queensland 4111, Australia}
	\author{Morgan M. Weston}
	\affiliation{Centre for Quantum Dynamics, Griffith University, Brisbane, Queensland 4111, Australia}
	\author{Helen M. Chrzanowski}
	\affiliation{Centre for Quantum Dynamics, Griffith University, Brisbane, Queensland 4111, Australia}
	\affiliation{Clarendon Laboratory, University of Oxford, Parks Road, Oxford OX1 3PU, UK}
	\author{Lynden K. Shalm}
	\affiliation{National Institute of Standards and Technology, 325 Broadway, Boulder, Colorado 80305, USA}
	\author{Varun B. Verma}
	\affiliation{National Institute of Standards and Technology, 325 Broadway, Boulder, Colorado 80305, USA}
	\author{Sae Woo Nam}
	\affiliation{National Institute of Standards and Technology, 325 Broadway, Boulder, Colorado 80305, USA}
	\author{Geoff J. Pryde}
	\email{g.pryde@griffith.edu.au}
	\affiliation{Centre for Quantum Dynamics, Griffith University, Brisbane, Queensland 4111, Australia}
	
\begin{abstract}
	 Interferometric phase measurement is widely used to precisely determine quantities such as length, speed, and material properties~\cite{caves81,book_wiseman_2009,giovannetti11}. Without quantum correlations, the best phase sensitivity $\Delta \varphi$ achievable using $n$ photons is the shot noise limit (SNL), $\Delta\varphi = 1/\sqrt{n}$. Quantum-enhanced metrology promises better sensitivity, but despite theoretical proposals stretching back decades~\cite{giovannetti11,rev_demkowicz15}, no measurement using photonic (i.e. definite photon number) quantum states has truly surpassed the SNL. Rather, all such demonstrations --- by discounting photon loss, detector inefficiency, or other imperfections --- have considered only a subset of the photons used. Here, we use an ultra-high efficiency photon source and detectors to perform unconditional entanglement-enhanced photonic interferometry. Sampling a birefringent phase shift, we demonstrate precision beyond the SNL without artificially correcting our results for loss and imperfections. Our results enable quantum-enhanced phase measurements at low photon flux and open the door to the next generation of optical quantum metrology advances.
 \end{abstract}
\maketitle 
It has been known for several decades that probing with various optical quantum states can achieve phase super-sensitivity, i.e measurement of the phase with an uncertainty below the SNL: ~\cite{giovannetti11,rev_demkowicz15}. It has been shown theoretically that multi-photon entangled states, such as NOON states, may achieve super-sensitivity and can, in principle, saturate the Heisenberg limit, the ultimate bound on sensitivity~\cite{dowling08,giovannetti11,rev_demkowicz15}. For this reason, they are of great interest for maximising the information that can be collected per photon, which is useful for investigating sensitive samples~\cite{wolfgramm13}. NOON states are superpositions of $N$ photons across two arms of an interferometer, each of which is a single optical mode: $\ket{\Psi_{NOON}}=1/\sqrt{2}(\ket{N}\ket{0}+\ket{0}\ket{N})$. We use the term photonic to refer to states like this, because they possess definite photon number, and these photons are counted in detection. By contrast, we exclude from term ``photonic'' schemes using states of indefinite photon number and continuous wave-like measurement, such as squeezed states and homodyne detection.  Such techniques have genuinely beaten the SNL, e.g.\ refs~\cite{yonezawa12,aasi13}, but work over narrow bandwidths and cannot directly achieve the theoretical maximal sensitivity per resource. The key feature of NOON and similar photonic states~\cite{xiang13} is that they produce interference fringes that oscillate faster any classical interference pattern, a feature called phase super-resolution~\cite{resch07}. Super-resolution interference experiments have been reported using two-\cite{ou90,rarity90,fonseca99,eisenberg05}, three-\cite{mitchell04}, four-~\cite{walther04,nagata07}, six-~\cite{resch07,xiang13} and eight-~\cite{gao10,wang16} photon states.  

Super-resolution, however, is not enough by itself to surpass the SNL~\cite{resch07,okamoto08,datta11}: a high interference fringe visibility, and high transmission and detection efficiency are also required---they must exceed the threshold at which these imperfections cancel the quantum advantage. For imperfect NOON state interferometry, a handy estimate of this threshold  was introduced in Ref.~\cite{resch07}: a genuine quantum advantage requires the interference visibility $v$ and the combined single-photon transmission and detection efficiency $\eta$ to satisfy  $\eta^N v^2 N > 1$. (For precise evaluation of the potential for super-sensitivity, the Fisher information can be used to analyse experimental schemes and data~\cite{datta11}, as described below.) 

Here we performed the first phase sensing experiment with $N=2$ photon NOON states that unconditionally demonstrates phase uncertainty below the SNL. This result was enabled by construction of a spontaneous parametric downconversion (SPDC) source~\cite{weston16}, that was optimised for high photon transmission and required no spectral filtering in order to achieve high quantum interference visibility $v$. This allowed us to fully exploit the benefits of high-efficiency detector configuration~\cite{marsili13}, yielding an ultra-high heralding efficiency~\cite{klyshko80}---equivalent to the single-photon efficiency $\eta$. We emphasise that high-efficiency ($95\%$) detection at $1550~\textrm{nm}$ wavelength has been available since 2008~\cite{lita08}, but photonic metrology beyond the SNL has not yet been achieved. This is because the capability to reach the required quantum interference and efficiency simultaneously requires a source capable of producing photons of exceptional spatial and spectral purity without filtering. 

Unlike previous experiments, our measurement apparatus does not require post-selection to achieve phase uncertainty below that achievable  in an  ideal, lossless classical interferometer. For our experimental apparatus, we expected $v\approx0.98$ and symmetrical interferometer arm efficiencies $\eta\approx0.8$ (which includes the detector efficiency), resulting in $\eta^N v^2 N \approx 1.23$. Thus, we anticipated a violation of the SNL, which we tested in two experiments described below.

For fair comparison with the SNL, an accurate accounting of resources is required. In the archetypal NOON-state phase sensing protocol, preparation and use of an $N$-photon NOON state constitutes a trial. In the ideal case, each trial leads to a detection event at the output of the interferometer.
Since each trial gives only a little information about the phase, a number of such trials may be performed. In our work, two-photon NOON states were generated probabilistically at random times by the SPDC source. Each detection event (i.e.\ any combination of detector registrations) represented a \textit{recorded} trial. We counted $k$ such detection events to complete the protocol. However, due to imperfect transmission and detection efficiency $\eta$, some NOON states did not lead to detections. Furthermore, due to higher-order SPDC events (the occasional simultaneous emission of $4,6,\ldots$ photons), the resources equivalent to multiple ($2,3,\ldots$) trials were overlapped in time and could not be distinguished by our non-photon-number-resolving detectors. Therefore, the actual number of trials (i.e.\ the number of photon pairs passing the phase shift), $\tilde{k}$, was larger than the number of recorded trials. Because the ideal classical scheme is assumed to be lossless and to use all resources passing the phase shift, it must be attributed an effective number of resources $n=N\tilde{k} = 2\tilde{k}$. This makes the SNL harder to beat. For the loss and downconversion parameters of our experiment, the worst-case estimate, based on the lowest possible value for the overall experimental efficiency, we determined $\tilde{k}/k=1.048125$.

\begin{figure}[t!]
	\centering
	\includegraphics[width=86mm]{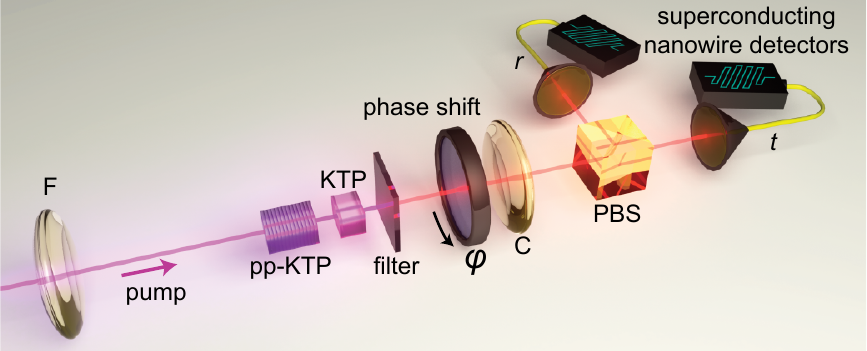}
	\caption[Metrology experimental set-up]{{\bf Experimental setup for the $N=2$ NOON state optical interferometer.} Laser pulses of $\sim 170$~fs duration, and centred at $775~\mathrm{nm}$, pump a $2~\mathrm{mm}$ periodically poled KTP (pp-KTP) crystal, phase-matched for type-II collinear, group-velocity-matched, degenerate down-conversion at $1550~\mathrm{nm}$. A compensation crystal ($1~\mathrm{mm}$ KTP) was used to compensate the temporal walk-off between signal and idler photons. Pump and collection spatial modes were set by F (focusing) and C (collimating) lenses correspondingly.  A silicon filter, AR-coated for $1550~\mathrm{nm}$, was used to block the pump beam. Transmitted ($t$) and reflected ($r$) modes of the interferometer, corresponding to H and V polarisations respectively, were separated by a polarising beam-splitter (PBS)  and then coupled into single mode fibres and sent to SNSPDs. A HWP mounted in an automated rotation stage is used to implement a controlled phase rotation, $\varphi$.}
	\label{fig:MetEXPsetup}
\end{figure}

Our experimental scheme is shown in Figure~1. We used collinear type-II parametric down-conversion, producing degenerate $1550~\mathrm{nm}$ photon pairs~\cite{weston16}. Careful design and implementation of the source's output mode structure allowed us to achieve high fiber-coupling efficiency and state-of-the-art superconducting nanowire single photon detectors (SNSPDs)~\cite{marsili13} provided high detection efficiency.  The down-conversion process generates two photons in the $\ket{1}_H\ket{1}_V$ polarisation state ($H\equiv$ horizontal; $V\equiv$ vertical) in the same spatial mode, which can be written as the NOON-polarisation state $\ket{\Psi}=\frac{1}{\sqrt{2}}(\ket{2}_L\ket{0}_R+\ket{0}_L\ket{2}_R)$. These right- (R) and left-circular (L) polarisation modes constituted the two arms of the interferometer. A half-wave plate (HWP) set at an angle $\varphi/4$ relative to its optic axis was used to implement the birefringent phase shift $\varphi$ between the arms. A common misconception about two-photon NOON states generated from SPDC is that the same phase sensitivity can be usefully achieved by using a pump photon (at half the wavelength) instead of the two-photon entangled state. However, this is clearly not correct for sensing in any material with dispersion.

After the phase shift, the modes were interfered on a polarising beam-splitter (PBS) and the output counting statistics were detected with SNSPDs and analysed with coincidence or time-tag logic. The output signal consisted of three possible types of detection outcomes: $C_{11}$, a coincidence detection between both output modes; $C_{20}$, a detection occurring only in the transmitted output mode; and $C_{02}$, a detection occurring only in the reflected output mode. The numbers of each type of detection in a time period $\tau$ were, respectively, $c_{11}(\varphi)$, $c_{20}(\varphi)$ and $c_{02}(\varphi)$. 

\begin{figure}[t!]
	\includegraphics[width=86mm]{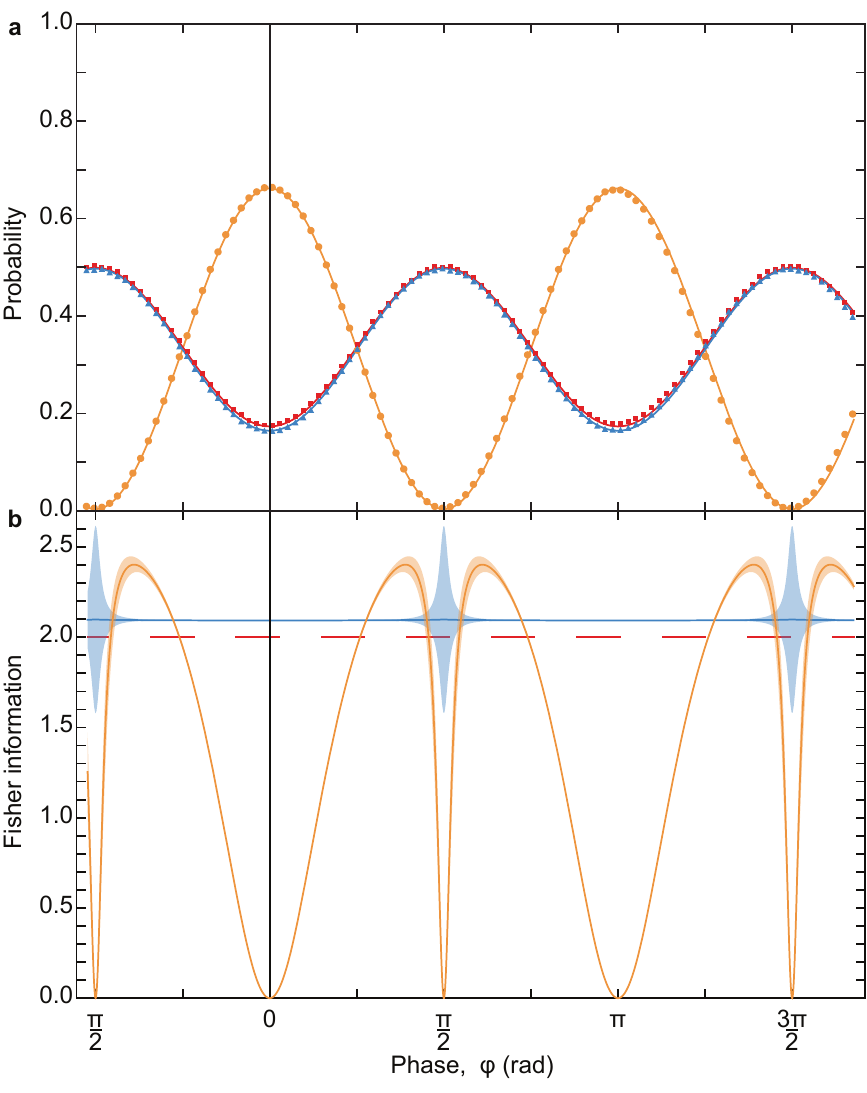}
	\caption[Experimentally measured fringes and Fisher information]{{\bf Experimentally measured output detection probability and the corresponding Fisher information.} \textbf{a}, Yellow circles, blue triangles and red squares represent our experimentally determined $p_{11}(\varphi)$, $p_{20}(\varphi)$ and $p_{02}(\varphi)$ probabilities, respectively. Error bars are smaller than the markers. Lines represent a theoretical model with corresponding transmission and interference visibility parameters. \textbf{b}, The yellow curve represents the Fisher information per recorded trial determined from the probability fringes, $p_{11}(\varphi),~p_{20}(\varphi)$ and $p_{02}(\varphi)$, as a function of the unknown phase $\varphi$. The dashed red line represents the Fisher information (per recorded trial) at the SNL, while the solid blue line represents the SNL Fisher information taking into account the inefficiency and multi-photon emission---see text and Methods for details. Shaded areas correspond to the $95\%$  confidence region, derived from the uncertainty in the fit parameters.}
	\label{fig:Metfringes}
\end{figure}
\begin{figure}[htb!]
	\centering
	\includegraphics[width=86mm]{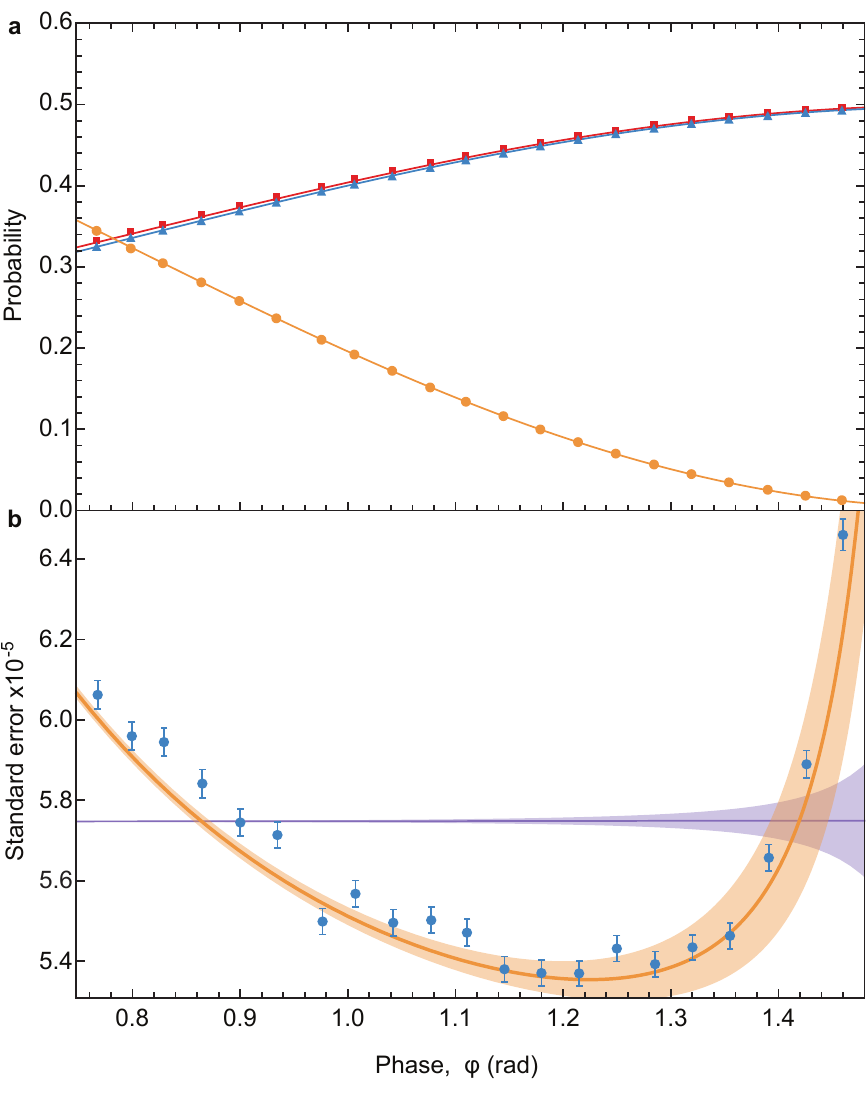}
	
	\caption[Experimentally measured phase estimate and phase uncertainty]{{\bf Experimentally measured phase estimate and phase uncertainty.} \textbf{a}, Lines represent theoretically modelled $p_{11}(\varphi),~p_{20}(\varphi)$ and $p_{02}(\varphi)$ probabilities from Fig.2{\bf a}. Yellow circles, blue triangles and red squares, correspond to average $P_{11}$, $P_{20}$, and $P_{02}$ probabilities, correspondingly, calculated from $145200000$ detection events per angle of HWP. Their positions on horizontal axis represent the measured phase estimate---see text for details. Error bars are smaller than the markers. \textbf{b}, Data points correspond to the standard deviation of the mean of corresponding phase measurements of {\bf a}. The purple line corresponds to standard deviation of the mean at the SNL, adjusted for correct number of resources $\tilde{N}$. The yellow line corresponds to the expected standard deviation of the mean, calculated from the Fisher information from Fig.~2{\bf b}. The shaded areas correspond to $95\%$  confidence regions, derived from the uncertainty in the fit parameters. The uncertainty in standard deviation of the mean was determined via the standard bootstrapping technique~\cite{book_davison_1997}, similarly to Ref.~\cite{higgins07}.}
	\label{fig:EXPerrors}
	
\end{figure}

In order to test and calibrate our setup we first measured interference fringes.
Detection events ($\approx250000$ per phase value) were collected for a fixed amount of time for various $\varphi \in [0,2\pi)$.
We observed an interference visibility of  $(98.9\pm0.02)\%$, calculated from fitting to the $c_{11}(\varphi)$ detection fringe. The transmissions of the reflected and transmitted outputs of the interferometer were measured to be $\eta_r=(79.41\pm0.09)\%$ and $\eta_t=(80.26\pm0.09)\%$, calculated from $c_{11}(0)/(c_{11}(0)+c_{20}(0))$  and $c_{11}(0)/(c_{11}(0)+c_{02}(0))$ ratios, respectively (a slight variation of transmission was observed when HWP was rotated, see Methods for details). Calculated transmissions include all the loss in the setup and the non-unit detection efficiency of SNSPDs. Probability fringes  $p_{11}(\varphi)$, $p_{20}(\varphi)$ and  $p_{02}(\varphi)$ were then obtained by fitting detection signals, $c_{i}(\varphi)$, $i \in \{11,20,02\}$, which were appropriately normalised for each phase value. We used the Fisher information per recorded trial,  $\mathcal{F}=\sum_{i}\left(\frac{\partial\ln{p_{i}}}{\partial\varphi} \right)^{2}p_i$ where $i\in \{11,20,02\}$, to quantify the phase sensitivity of our phase measurement setup~\cite{giovannetti11}. Our results (Fig.~2b) show a clear violation, for a range of phase values $\varphi$, of the adjusted SNL bound that takes into account the information in unrecorded trials: $\mathcal{F}_{SNL}=N\tilde{k}/k=2.09625$.

For the second experiment, we performed phase sensing for individual settings of the phase shifter within the range where we expected to beat the SNL. At each setting, time-tag hardware was used to acquire detection events corresponding to $k=10000$ trials. From the distribution of $C_i$ events ($i \in \{11,20,02\}$), corresponding $P_i$ probabilities were obtained by normalisation. 
This set of three probabilities corresponds to a single phase estimate value $\varphi^\textrm{\scriptsize est}$. In practice, finding this estimate required to minimise the squared difference between the measured probabilities and their corresponding calibration curves, $p_i(\varphi)$ (Fig.~2a). 
The phase search range was restricted to $\varphi^\textrm{\scriptsize est}\in[0,\pi/2]$. 
This process was repeated for $s=14520$ samples (for each angle of the HWP), determining $\varphi^\textrm{\scriptsize est}_j$ for each sample $j$. The mean and standard deviation of the mean (standard error of the mean) $\{\varphi^\textrm{\scriptsize est}_j\}$ are shown in Fig.~3.
This measurement procedure was repeated for a range of phase values around the region of interest.
When compared to the standard deviation of the mean ($=1/\sqrt{n^\textrm{\scriptsize tot}}$)  that is achievable with $n^\textrm{\scriptsize tot}=ns=N\tilde{k}s=304375500$ classical resources (adjusted for loss and higher order terms, as before) per data point, our results show a clear advantage of our quantum approach.

In conclusion, we demonstrated unconditional violation of the SNL in photonic phase sensing. We recover the estimate of the phase shift applied to the mode, and its corresponding standard deviation of the mean, directly from our measurement data without additional adjustments.  Moreover, all the parameters necessary for the calculation of the SNL of our measurement apparatus, such as circuit loss, pair and multi-pair generation probability, were calculated directly from our measurement data. Our results are not only of fundamental interest, but are also directly applicable to a phase measurement scenario where low photon flux is required, such as measurement of light-sensitive materials~\cite{wolfgramm13}. We note that the $N=2$ NOON state is also the $N=2$ Holland-Burnett state; thus, we anticipate that our technique can be extended, using loss-tolerant approaches~\cite{matthews16} and number-resolving detectors~\cite{harder16}, to perform sub-SNL sensing with significantly higher photon numbers in the near future. It can also be applied to multipass metrology protocols~\cite{higgins07}.

\section*{Methods}
\subsection*{Photon source and characterisation}
High heralding efficiency (Klyshko efficiency~\cite{klyshko80}) and high interference visibility were essential to the success of this demonstration. These features were achieved by using a spontaneous parametric down-conversion (SPDC) source that operated at the group velocity matching conditions to generate frequency uncorrelated photon pairs, removing the need for spectral filtering that is typically required in conventional SPDC sources. The heralded single photon source was pumped by a mode-locked Ti:Sapphire laser with 81 MHz repetition rate, $775~\mathrm{nm}$ wavelength and $6~\mathrm{nm}$ FWHM bandwidth. SPDC from a nonlinear periodically poled KTP crystal (pp-KTP, poling period $46.20~\mathrm{\mu m}$) phase matched for type-II collinear operation produced degenerate photon pairs centred at $1550~\mathrm{nm}$ wavelength and with $\approx 15~\mathrm{nm}$ FWHM bandwidth. We used a pump waist size of $170~\mathrm{\mu m}$ and a signal and idler collection waist size of $50~\mathrm{\mu m}$~\cite{weston16}. Together with the high efficiency SNSPDs~\cite{marsili13}, this allowed us to achieve symmetric heralding efficiencies of $(82\pm 2)\%$. 
An additional $1~\mathrm{mm}$ KTP crystal, with optic axis rotated by $90^{\circ}$ with respect to the optic axis of pp-KTP, was placed after the downconverter to compensate for the temporal walk-off (due to birefringence) in the $2~\mathrm{mm}$ pp-KTP crystal. Non-classical interference visibilities between the signal and idler photons were measured to be $(98.9\pm0.2)\%$, with no background subtraction.

\subsection*{Detection}
There is one superconducting nanowire single photon detector (SNSPD) coupled to each of the interferometer outputs. We record three types of events: $C_{11}$, a coincidence detection between both output modes; $C_{20}$, a detection occurring only in the transmitted output mode; and $C_{02}$, a detection occurring only in the reflected output mode. Since the SNSPDs cannot resolve photon number, the latter two events may arise from: the case where both photons in the two-photon NOON state went to the same detector; a case where one photon was lost and therefore only one photon is recorded; or the case of a dark count. We do not artificially remove signals from these latter two cases. We note that the rate of dark counts is small compared to the rate of real events.

\subsection*{Resource counting}
The total transmission (including detection efficiency) is easily determined when the phase shift is set to zero. However, we have also observed that the circuit transmission can slightly vary with the rotation of the HWP that implements the phase shift. In order to verify transmission at every value of $\varphi$ deviates only slightly from this, we build a theoretical model of the interferometer, which includes an imperfect interference on the PBS and loss in two of the output modes. Then, at each rotation angle of the HWP we find transmission of both output modes ($\eta_r(\varphi)$ and $\eta_t(\varphi)$) by finding the closest theoretical match to the measured data with least squares minimisation. We confirm that the transmission variation stays below $\approx1\%$. Moreover, the standard deviation in total number of detection events per point in Fig.~2, which were acquired over a fixed amount of time, was measured to be $\approx0.2\%$, confirming that the variation of our setup parameters was very small.

In order to estimate the true number of resources we use, we calculated the probability of having at least one photon transmitted through the system. For the three possible outcomes, this is given by
\begin{eqnarray}
\label{eq:minloss}
\eta_{11}(\varphi)=1-(1-\eta_t(\varphi))(1-\eta_r(\varphi)),\\ \nonumber
\eta_{20}(\varphi)=1-(1-\eta_t(\varphi))(1-\eta_t(\varphi)),\\ 
\eta_{02}(\varphi)=1-(1-\eta_r(\varphi))(1-\eta_r(\varphi)). \nonumber
\end{eqnarray}
The lowest efficiency $\eta_{min}=\min_{\varphi,j}(\eta_j(\varphi))$, for $j\in\{11,20,02\}$ and $\varphi\in[0,2\pi)$, observed over entire range of $\varphi$ out of the three outcomes was $\eta_{min}\approx95.56\%$. We chose the lowest efficiency $\eta_{min}(\varphi)=\min_{j}(\eta_j(\varphi))$, calculated at each angle of the HWP (i.e.\ at each $\varphi$) in order to calculate the number of trials accordingly. This results in an overestimate of the number of photons used in our experiment---and therefore used to calculate the effective SNL---setting a higher SNL bound than was actually the case, and making it harder to violate the SNL. The uncertainty in the estimated transmission and the SNL bound was calculated by error propagation of the uncertainty in $c_i$ through the theoretical model of the interferometer. We have observed very strong dependence of the estimated parameters on $c_{11}(\varphi)$ near the $\varphi=\pi/2+k\pi$, $k\in\mathbb{Z}$, resulting in an increased parameter uncertainty in those regions.

Another important source of additional resources that are not directly detected in the protocol is the possible emission of multiple photon pairs in the SPDC process. For the pump power of $\approx\mathrm{100~mW}$ used in our experiment, we measured this probability to be $\xi=(0.1550\pm0.0004)\%$, relative to the probability of generating a single pair. This was done by comparing the number of photon detection events within a fixed time window with the actual number of laser pulses in that window, acquired with a separate trigger signal. We determined $\xi$ by carefully determining the single pair generation probability,
with loss taken into account.

The actual number of trials (pairs of photons generated, for $N=2$) $\tilde{k}$ is related to the number of recorded trials $k$ by
\begin{eqnarray}
\label{eq:recourcen}
\tilde{k}(\varphi)=\frac{k(1+\xi)}{\eta_{min}(\varphi)}. 
\end{eqnarray}
Equivalently, the actual average number of resources per recorded trial is 
\begin{eqnarray}
\frac{N\tilde{k}}{k}=\frac{2(1+\xi)}{\eta_{min}(\varphi)},
\end{eqnarray}
for $N=2$.


\begin{thebibliography}{29}%
	\makeatletter
	\providecommand \@ifxundefined [1]{%
		\@ifx{#1\undefined}
	}%
	\providecommand \@ifnum [1]{%
		\ifnum #1\expandafter \@firstoftwo
		\else \expandafter \@secondoftwo
		\fi
	}%
	\providecommand \@ifx [1]{%
		\ifx #1\expandafter \@firstoftwo
		\else \expandafter \@secondoftwo
		\fi
	}%
	\providecommand \natexlab [1]{#1}%
	\providecommand \enquote  [1]{``#1''}%
	\providecommand \bibnamefont  [1]{#1}%
	\providecommand \bibfnamefont [1]{#1}%
	\providecommand \citenamefont [1]{#1}%
	\providecommand \href@noop [0]{\@secondoftwo}%
	\providecommand \href [0]{\begingroup \@sanitize@url \@href}%
	\providecommand \@href[1]{\@@startlink{#1}\@@href}%
	\providecommand \@@href[1]{\endgroup#1\@@endlink}%
	\providecommand \@sanitize@url [0]{\catcode `\\12\catcode `\$12\catcode
		`\&12\catcode `\#12\catcode `\^12\catcode `\_12\catcode `\%12\relax}%
	\providecommand \@@startlink[1]{}%
	\providecommand \@@endlink[0]{}%
	\providecommand \url  [0]{\begingroup\@sanitize@url \@url }%
	\providecommand \@url [1]{\endgroup\@href {#1}{\urlprefix }}%
	\providecommand \urlprefix  [0]{URL }%
	\providecommand \Eprint [0]{\href }%
	\providecommand \doibase [0]{http://dx.doi.org/}%
	\providecommand \selectlanguage [0]{\@gobble}%
	\providecommand \bibinfo  [0]{\@secondoftwo}%
	\providecommand \bibfield  [0]{\@secondoftwo}%
	\providecommand \translation [1]{[#1]}%
	\providecommand \BibitemOpen [0]{}%
	\providecommand \bibitemStop [0]{}%
	\providecommand \bibitemNoStop [0]{.\EOS\space}%
	\providecommand \EOS [0]{\spacefactor3000\relax}%
	\providecommand \BibitemShut  [1]{\csname bibitem#1\endcsname}%
	\let\auto@bib@innerbib\@empty
	\bibitem [{\citenamefont {Caves}(1981)}]{caves81}%
	\BibitemOpen
	\bibfield  {author} {\bibinfo {author} {\bibfnamefont {C.~M.}\ \bibnamefont
			{Caves}},\ }\href {\doibase 10.1103/PhysRevD.23.1693} {\bibfield  {journal}
		{\bibinfo  {journal} {Phys. Rev. D}\ }\textbf {\bibinfo {volume} {23}},\
		\bibinfo {pages} {1693} (\bibinfo {year} {1981})}\BibitemShut {NoStop}%
	\bibitem [{\citenamefont {Wiseman}\ and\ \citenamefont
		{Milburn}(2009)}]{book_wiseman_2009}%
	\BibitemOpen
	\bibfield  {author} {\bibinfo {author} {\bibfnamefont {H.~M.}\ \bibnamefont
			{Wiseman}}\ and\ \bibinfo {author} {\bibfnamefont {G.~J.}\ \bibnamefont
			{Milburn}},\ }\href@noop {} {\emph {\bibinfo {title} {Quantum Measurement and
				Control}}}\ (\bibinfo  {publisher} {Cambridge Univ. Press},\ \bibinfo {year}
	{2009})\BibitemShut {NoStop}%
	\bibitem [{\citenamefont {Giovannetti}\ \emph {et~al.}(2011)\citenamefont
		{Giovannetti}, \citenamefont {Lloyd},\ and\ \citenamefont
		{Maccone}}]{giovannetti11}%
	\BibitemOpen
	\bibfield  {author} {\bibinfo {author} {\bibfnamefont {V.}~\bibnamefont
			{Giovannetti}}, \bibinfo {author} {\bibfnamefont {S.}~\bibnamefont {Lloyd}},
		\ and\ \bibinfo {author} {\bibfnamefont {L.}~\bibnamefont {Maccone}},\ }\href
	{\doibase 10.1038/NPHOTON.2011.35} {\bibfield  {journal} {\bibinfo  {journal}
			{Nat. Photon.}\ }\textbf {\bibinfo {volume} {5}},\ \bibinfo {pages} {222}
		(\bibinfo {year} {2011})}\BibitemShut {NoStop}%
	\bibitem [{\citenamefont {Demkowicz-Dobrza{\'n}ski}\ \emph
		{et~al.}(2015)\citenamefont {Demkowicz-Dobrza{\'n}ski}, \citenamefont
		{Jarzyna},\ and\ \citenamefont {Ko{\l}ody{\'n}ski}}]{rev_demkowicz15}%
	\BibitemOpen
	\bibfield  {author} {\bibinfo {author} {\bibfnamefont {R.}~\bibnamefont
			{Demkowicz-Dobrza{\'n}ski}}, \bibinfo {author} {\bibfnamefont
			{M.}~\bibnamefont {Jarzyna}}, \ and\ \bibinfo {author} {\bibfnamefont
			{J.}~\bibnamefont {Ko{\l}ody{\'n}ski}},\ }\href {\doibase
		10.1016/bs.po.2015.02.003} {\bibfield  {journal} {\bibinfo  {journal} {Prog.
				Opt.}\ }\textbf {\bibinfo {volume} {60}},\ \bibinfo {pages} {345} (\bibinfo
		{year} {2015})}\BibitemShut {NoStop}%
	\bibitem [{\citenamefont {Dowling}(2008)}]{dowling08}%
	\BibitemOpen
	\bibfield  {author} {\bibinfo {author} {\bibfnamefont {J.~P.}\ \bibnamefont
			{Dowling}},\ }\href {\doibase 10.1080/00107510802091298} {\bibfield
		{journal} {\bibinfo  {journal} {Contemp. Phys.}\ }\textbf {\bibinfo {volume}
			{49}},\ \bibinfo {pages} {125} (\bibinfo {year} {2008})}\BibitemShut
	{NoStop}%
	\bibitem [{\citenamefont {Wolfgramm}\ \emph {et~al.}(2013)\citenamefont
		{Wolfgramm}, \citenamefont {Vitelli}, \citenamefont {Beduini}, \citenamefont
		{Godbout},\ and\ \citenamefont {Mitchell}}]{wolfgramm13}%
	\BibitemOpen
	\bibfield  {author} {\bibinfo {author} {\bibfnamefont {F.}~\bibnamefont
			{Wolfgramm}}, \bibinfo {author} {\bibfnamefont {C.}~\bibnamefont {Vitelli}},
		\bibinfo {author} {\bibfnamefont {F.~A.}\ \bibnamefont {Beduini}}, \bibinfo
		{author} {\bibfnamefont {N.}~\bibnamefont {Godbout}}, \ and\ \bibinfo
		{author} {\bibfnamefont {M.~W.}\ \bibnamefont {Mitchell}},\ }\href
	{http://dx.doi.org/10.1038/nphoton.2012.300} {\bibfield  {journal} {\bibinfo
			{journal} {Nat. Photon.}\ }\textbf {\bibinfo {volume} {7}},\ \bibinfo {pages}
		{28} (\bibinfo {year} {2013})}\BibitemShut {NoStop}%
	\bibitem [{\citenamefont {Yonezawa}\ \emph {et~al.}(2012)\citenamefont
		{Yonezawa}, \citenamefont {Nakane}, \citenamefont {Wheatley}, \citenamefont
		{Iwasawa}, \citenamefont {Takeda}, \citenamefont {Arao}, \citenamefont
		{Ohki}, \citenamefont {Tsumura}, \citenamefont {Berry}, \citenamefont
		{Ralph}, \citenamefont {Wiseman}, \citenamefont {Huntington},\ and\
		\citenamefont {Furusawa}}]{yonezawa12}%
	\BibitemOpen
	\bibfield  {author} {\bibinfo {author} {\bibfnamefont {H.}~\bibnamefont
			{Yonezawa}}, \bibinfo {author} {\bibfnamefont {D.}~\bibnamefont {Nakane}},
		\bibinfo {author} {\bibfnamefont {T.~A.}\ \bibnamefont {Wheatley}}, \bibinfo
		{author} {\bibfnamefont {K.}~\bibnamefont {Iwasawa}}, \bibinfo {author}
		{\bibfnamefont {S.}~\bibnamefont {Takeda}}, \bibinfo {author} {\bibfnamefont
			{H.}~\bibnamefont {Arao}}, \bibinfo {author} {\bibfnamefont {K.}~\bibnamefont
			{Ohki}}, \bibinfo {author} {\bibfnamefont {K.}~\bibnamefont {Tsumura}},
		\bibinfo {author} {\bibfnamefont {D.~W.}\ \bibnamefont {Berry}}, \bibinfo
		{author} {\bibfnamefont {T.~C.}\ \bibnamefont {Ralph}}, \bibinfo {author}
		{\bibfnamefont {H.~M.}\ \bibnamefont {Wiseman}}, \bibinfo {author}
		{\bibfnamefont {E.~H.}\ \bibnamefont {Huntington}}, \ and\ \bibinfo {author}
		{\bibfnamefont {A.}~\bibnamefont {Furusawa}},\ }\href {\doibase
		10.1126/science.1225258} {\bibfield  {journal} {\bibinfo  {journal}
			{Science}\ }\textbf {\bibinfo {volume} {337}},\ \bibinfo {pages} {1514}
		(\bibinfo {year} {2012})}\BibitemShut {NoStop}%
	\bibitem [{\citenamefont {Aasi}\ \emph {et~al.}(2013)\citenamefont {Aasi},
		\citenamefont {Abadie}, \citenamefont {Abbott}, \citenamefont {Abbott},
		\citenamefont {Abbott}, \citenamefont {Abernathy}, \citenamefont {Adams},
		\citenamefont {Adams}, \citenamefont {Addesso}, \citenamefont {Adhikari}
		\emph {et~al.}}]{aasi13}%
	\BibitemOpen
	\bibfield  {author} {\bibinfo {author} {\bibfnamefont {J.}~\bibnamefont
			{Aasi}}, \bibinfo {author} {\bibfnamefont {J.}~\bibnamefont {Abadie}},
		\bibinfo {author} {\bibfnamefont {B.}~\bibnamefont {Abbott}}, \bibinfo
		{author} {\bibfnamefont {R.}~\bibnamefont {Abbott}}, \bibinfo {author}
		{\bibfnamefont {T.}~\bibnamefont {Abbott}}, \bibinfo {author} {\bibfnamefont
			{M.}~\bibnamefont {Abernathy}}, \bibinfo {author} {\bibfnamefont
			{C.}~\bibnamefont {Adams}}, \bibinfo {author} {\bibfnamefont
			{T.}~\bibnamefont {Adams}}, \bibinfo {author} {\bibfnamefont
			{P.}~\bibnamefont {Addesso}}, \bibinfo {author} {\bibfnamefont
			{R.}~\bibnamefont {Adhikari}},  \emph {et~al.},\ }\href {\doibase
		10.1038/nphoton.2013.177} {\bibfield  {journal} {\bibinfo  {journal} {Nat.
				Photon.}\ }\textbf {\bibinfo {volume} {7}},\ \bibinfo {pages} {613} (\bibinfo
		{year} {2013})}\BibitemShut {NoStop}%
	\bibitem [{\citenamefont {Xiang}\ \emph {et~al.}(2013)\citenamefont {Xiang},
		\citenamefont {Hofmann},\ and\ \citenamefont {Pryde}}]{xiang13}%
	\BibitemOpen
	\bibfield  {author} {\bibinfo {author} {\bibfnamefont {G.}~\bibnamefont
			{Xiang}}, \bibinfo {author} {\bibfnamefont {H.}~\bibnamefont {Hofmann}}, \
		and\ \bibinfo {author} {\bibfnamefont {G.~J.}\ \bibnamefont {Pryde}},\ }\href
	{\doibase 10.1038/srep02684} {\bibfield  {journal} {\bibinfo  {journal} {Sci.
				Rep.}\ }\textbf {\bibinfo {volume} {3}},\ \bibinfo {pages} {2684} (\bibinfo
		{year} {2013})}\BibitemShut {NoStop}%
	\bibitem [{\citenamefont {Resch}\ \emph {et~al.}(2007)\citenamefont {Resch},
		\citenamefont {Pregnell}, \citenamefont {Prevedel}, \citenamefont
		{Gilchrist}, \citenamefont {Pryde}, \citenamefont {O'Brien},\ and\
		\citenamefont {White}}]{resch07}%
	\BibitemOpen
	\bibfield  {author} {\bibinfo {author} {\bibfnamefont {K.~J.}\ \bibnamefont
			{Resch}}, \bibinfo {author} {\bibfnamefont {K.~L.}\ \bibnamefont {Pregnell}},
		\bibinfo {author} {\bibfnamefont {R.}~\bibnamefont {Prevedel}}, \bibinfo
		{author} {\bibfnamefont {A.}~\bibnamefont {Gilchrist}}, \bibinfo {author}
		{\bibfnamefont {G.~J.}\ \bibnamefont {Pryde}}, \bibinfo {author}
		{\bibfnamefont {J.~L.}\ \bibnamefont {O'Brien}}, \ and\ \bibinfo {author}
		{\bibfnamefont {A.~G.}\ \bibnamefont {White}},\ }\href {\doibase
		10.1103/PhysRevLett.98.223601} {\bibfield  {journal} {\bibinfo  {journal}
			{Phys. Rev. Lett.}\ }\textbf {\bibinfo {volume} {98}},\ \bibinfo {pages}
		{223601} (\bibinfo {year} {2007})}\BibitemShut {NoStop}%
	\bibitem [{\citenamefont {Ou}\ \emph {et~al.}(1990)\citenamefont {Ou},
		\citenamefont {Zou}, \citenamefont {Wang},\ and\ \citenamefont
		{Mandel}}]{ou90}%
	\BibitemOpen
	\bibfield  {author} {\bibinfo {author} {\bibfnamefont {Z.~Y.}\ \bibnamefont
			{Ou}}, \bibinfo {author} {\bibfnamefont {X.~Y.}\ \bibnamefont {Zou}},
		\bibinfo {author} {\bibfnamefont {L.~J.}\ \bibnamefont {Wang}}, \ and\
		\bibinfo {author} {\bibfnamefont {L.}~\bibnamefont {Mandel}},\ }\href
	{\doibase 10.1103/PhysRevA.42.2957} {\bibfield  {journal} {\bibinfo
			{journal} {Phys. Rev. A}\ }\textbf {\bibinfo {volume} {42}},\ \bibinfo
		{pages} {2957} (\bibinfo {year} {1990})}\BibitemShut {NoStop}%
	\bibitem [{\citenamefont {Rarity}\ \emph {et~al.}(1990)\citenamefont {Rarity},
		\citenamefont {Tapster}, \citenamefont {Jakeman}, \citenamefont {Larchuk},
		\citenamefont {Campos}, \citenamefont {Teich},\ and\ \citenamefont
		{Saleh}}]{rarity90}%
	\BibitemOpen
	\bibfield  {author} {\bibinfo {author} {\bibfnamefont {J.~G.}\ \bibnamefont
			{Rarity}}, \bibinfo {author} {\bibfnamefont {P.~R.}\ \bibnamefont {Tapster}},
		\bibinfo {author} {\bibfnamefont {E.}~\bibnamefont {Jakeman}}, \bibinfo
		{author} {\bibfnamefont {T.}~\bibnamefont {Larchuk}}, \bibinfo {author}
		{\bibfnamefont {R.~A.}\ \bibnamefont {Campos}}, \bibinfo {author}
		{\bibfnamefont {M.~C.}\ \bibnamefont {Teich}}, \ and\ \bibinfo {author}
		{\bibfnamefont {B.~E.~A.}\ \bibnamefont {Saleh}},\ }\href {\doibase
		10.1103/PhysRevLett.65.1348} {\bibfield  {journal} {\bibinfo  {journal}
			{Phys. Rev. Lett.}\ }\textbf {\bibinfo {volume} {65}},\ \bibinfo {pages}
		{1348} (\bibinfo {year} {1990})}\BibitemShut {NoStop}%
	\bibitem [{\citenamefont {Fonseca}\ \emph {et~al.}(1999)\citenamefont
		{Fonseca}, \citenamefont {Monken},\ and\ \citenamefont
		{P\'adua}}]{fonseca99}%
	\BibitemOpen
	\bibfield  {author} {\bibinfo {author} {\bibfnamefont {E.~J.~S.}\
			\bibnamefont {Fonseca}}, \bibinfo {author} {\bibfnamefont {C.~H.}\
			\bibnamefont {Monken}}, \ and\ \bibinfo {author} {\bibfnamefont
			{S.}~\bibnamefont {P\'adua}},\ }\href {\doibase 10.1103/PhysRevLett.82.2868}
	{\bibfield  {journal} {\bibinfo  {journal} {Phys. Rev. Lett.}\ }\textbf
		{\bibinfo {volume} {82}},\ \bibinfo {pages} {2868} (\bibinfo {year}
		{1999})}\BibitemShut {NoStop}%
	\bibitem [{\citenamefont {Eisenberg}\ \emph {et~al.}(2005)\citenamefont
		{Eisenberg}, \citenamefont {Hodelin}, \citenamefont {Khoury},\ and\
		\citenamefont {Bouwmeester}}]{eisenberg05}%
	\BibitemOpen
	\bibfield  {author} {\bibinfo {author} {\bibfnamefont {H.~S.}\ \bibnamefont
			{Eisenberg}}, \bibinfo {author} {\bibfnamefont {J.~F.}\ \bibnamefont
			{Hodelin}}, \bibinfo {author} {\bibfnamefont {G.}~\bibnamefont {Khoury}}, \
		and\ \bibinfo {author} {\bibfnamefont {D.}~\bibnamefont {Bouwmeester}},\
	}\href {\doibase 10.1103/PhysRevLett.94.090502} {\bibfield  {journal}
		{\bibinfo  {journal} {Phys. Rev. Lett.}\ }\textbf {\bibinfo {volume} {94}},\
		\bibinfo {pages} {090502} (\bibinfo {year} {2005})}\BibitemShut {NoStop}%
	\bibitem [{\citenamefont {Mitchell}\ \emph {et~al.}(2004)\citenamefont
		{Mitchell}, \citenamefont {Lundeen},\ and\ \citenamefont
		{Steinberg}}]{mitchell04}%
	\BibitemOpen
	\bibfield  {author} {\bibinfo {author} {\bibfnamefont {M.~W.}\ \bibnamefont
			{Mitchell}}, \bibinfo {author} {\bibfnamefont {J.~S.}\ \bibnamefont
			{Lundeen}}, \ and\ \bibinfo {author} {\bibfnamefont {A.~M.}\ \bibnamefont
			{Steinberg}},\ }\href {\doibase 10.1038/nature02493} {\bibfield  {journal}
		{\bibinfo  {journal} {Nature}\ }\textbf {\bibinfo {volume} {429}},\ \bibinfo
		{pages} {161} (\bibinfo {year} {2004})}\BibitemShut {NoStop}%
	\bibitem [{\citenamefont {Walther}\ \emph {et~al.}(2004)\citenamefont
		{Walther}, \citenamefont {Pan}, \citenamefont {Aspelmeyer}, \citenamefont
		{Ursin}, \citenamefont {Gasparoni},\ and\ \citenamefont
		{Zeilinger}}]{walther04}%
	\BibitemOpen
	\bibfield  {author} {\bibinfo {author} {\bibfnamefont {P.}~\bibnamefont
			{Walther}}, \bibinfo {author} {\bibfnamefont {J.-W.}\ \bibnamefont {Pan}},
		\bibinfo {author} {\bibfnamefont {M.}~\bibnamefont {Aspelmeyer}}, \bibinfo
		{author} {\bibfnamefont {R.}~\bibnamefont {Ursin}}, \bibinfo {author}
		{\bibfnamefont {S.}~\bibnamefont {Gasparoni}}, \ and\ \bibinfo {author}
		{\bibfnamefont {A.}~\bibnamefont {Zeilinger}},\ }\href {\doibase
		10.1038/nature02552} {\bibfield  {journal} {\bibinfo  {journal} {Nature}\
		}\textbf {\bibinfo {volume} {429}},\ \bibinfo {pages} {158} (\bibinfo {year}
		{2004})}\BibitemShut {NoStop}%
	\bibitem [{\citenamefont {Nagata}\ \emph {et~al.}(2007)\citenamefont {Nagata},
		\citenamefont {Okamoto}, \citenamefont {O'Brien}, \citenamefont {Sasaki},\
		and\ \citenamefont {Takeuchi}}]{nagata07}%
	\BibitemOpen
	\bibfield  {author} {\bibinfo {author} {\bibfnamefont {T.}~\bibnamefont
			{Nagata}}, \bibinfo {author} {\bibfnamefont {R.}~\bibnamefont {Okamoto}},
		\bibinfo {author} {\bibfnamefont {J.~L.}\ \bibnamefont {O'Brien}}, \bibinfo
		{author} {\bibfnamefont {K.}~\bibnamefont {Sasaki}}, \ and\ \bibinfo {author}
		{\bibfnamefont {S.}~\bibnamefont {Takeuchi}},\ }\href {\doibase
		10.1126/science.1138007} {\bibfield  {journal} {\bibinfo  {journal}
			{Science}\ }\textbf {\bibinfo {volume} {316}},\ \bibinfo {pages} {726}
		(\bibinfo {year} {2007})}\BibitemShut {NoStop}%
	\bibitem [{\citenamefont {Gao}\ \emph {et~al.}(2010)\citenamefont {Gao},
		\citenamefont {Lu}, \citenamefont {Yao}, \citenamefont {Xu}, \citenamefont
		{G{\"u}hne}, \citenamefont {Goebel}, \citenamefont {Chen}, \citenamefont
		{Peng}, \citenamefont {Chen},\ and\ \citenamefont {Pan}}]{gao10}%
	\BibitemOpen
	\bibfield  {author} {\bibinfo {author} {\bibfnamefont {W.-B.}\ \bibnamefont
			{Gao}}, \bibinfo {author} {\bibfnamefont {C.-Y.}\ \bibnamefont {Lu}},
		\bibinfo {author} {\bibfnamefont {X.-C.}\ \bibnamefont {Yao}}, \bibinfo
		{author} {\bibfnamefont {P.}~\bibnamefont {Xu}}, \bibinfo {author}
		{\bibfnamefont {O.}~\bibnamefont {G{\"u}hne}}, \bibinfo {author}
		{\bibfnamefont {A.}~\bibnamefont {Goebel}}, \bibinfo {author} {\bibfnamefont
			{Y.-A.}\ \bibnamefont {Chen}}, \bibinfo {author} {\bibfnamefont {C.-Z.}\
			\bibnamefont {Peng}}, \bibinfo {author} {\bibfnamefont {Z.-B.}\ \bibnamefont
			{Chen}}, \ and\ \bibinfo {author} {\bibfnamefont {J.-W.}\ \bibnamefont
			{Pan}},\ }\href {\doibase 10.1038/nphys1603} {\bibfield  {journal} {\bibinfo
			{journal} {Nat. Phys.}\ }\textbf {\bibinfo {volume} {6}},\ \bibinfo {pages}
		{331} (\bibinfo {year} {2010})}\BibitemShut {NoStop}%
	\bibitem [{\citenamefont {Wang}\ \emph {et~al.}(2016)\citenamefont {Wang},
		\citenamefont {Chen}, \citenamefont {Li}, \citenamefont {Huang},
		\citenamefont {Liu}, \citenamefont {Chen}, \citenamefont {Luo}, \citenamefont
		{Su}, \citenamefont {Wu}, \citenamefont {Li}, \citenamefont {Lu},
		\citenamefont {Hu}, \citenamefont {Jiang}, \citenamefont {Peng},
		\citenamefont {Li}, \citenamefont {Liu}, \citenamefont {Chen}, \citenamefont
		{Lu},\ and\ \citenamefont {Pan}}]{wang16}%
	\BibitemOpen
	\bibfield  {author} {\bibinfo {author} {\bibfnamefont {X.-L.}\ \bibnamefont
			{Wang}}, \bibinfo {author} {\bibfnamefont {L.-K.}\ \bibnamefont {Chen}},
		\bibinfo {author} {\bibfnamefont {W.}~\bibnamefont {Li}}, \bibinfo {author}
		{\bibfnamefont {H.-L.}\ \bibnamefont {Huang}}, \bibinfo {author}
		{\bibfnamefont {C.}~\bibnamefont {Liu}}, \bibinfo {author} {\bibfnamefont
			{C.}~\bibnamefont {Chen}}, \bibinfo {author} {\bibfnamefont {Y.-H.}\
			\bibnamefont {Luo}}, \bibinfo {author} {\bibfnamefont {Z.-E.}\ \bibnamefont
			{Su}}, \bibinfo {author} {\bibfnamefont {D.}~\bibnamefont {Wu}}, \bibinfo
		{author} {\bibfnamefont {Z.-D.}\ \bibnamefont {Li}}, \bibinfo {author}
		{\bibfnamefont {H.}~\bibnamefont {Lu}}, \bibinfo {author} {\bibfnamefont
			{Y.}~\bibnamefont {Hu}}, \bibinfo {author} {\bibfnamefont {X.}~\bibnamefont
			{Jiang}}, \bibinfo {author} {\bibfnamefont {C.-Z.}\ \bibnamefont {Peng}},
		\bibinfo {author} {\bibfnamefont {L.}~\bibnamefont {Li}}, \bibinfo {author}
		{\bibfnamefont {N.-L.}\ \bibnamefont {Liu}}, \bibinfo {author} {\bibfnamefont
			{Y.-A.}\ \bibnamefont {Chen}}, \bibinfo {author} {\bibfnamefont {C.-Y.}\
			\bibnamefont {Lu}}, \ and\ \bibinfo {author} {\bibfnamefont {J.-W.}\
			\bibnamefont {Pan}},\ }\href {\doibase 10.1103/PhysRevLett.117.210502}
	{\bibfield  {journal} {\bibinfo  {journal} {Phys. Rev. Lett.}\ }\textbf
		{\bibinfo {volume} {117}},\ \bibinfo {pages} {210502} (\bibinfo {year}
		{2016})}\BibitemShut {NoStop}%
	\bibitem [{\citenamefont {Okamoto}\ \emph {et~al.}(2008)\citenamefont
		{Okamoto}, \citenamefont {Hofmann}, \citenamefont {Nagata}, \citenamefont
		{O'Brien}, \citenamefont {Sasaki},\ and\ \citenamefont
		{Takeuchi}}]{okamoto08}%
	\BibitemOpen
	\bibfield  {author} {\bibinfo {author} {\bibfnamefont {R.}~\bibnamefont
			{Okamoto}}, \bibinfo {author} {\bibfnamefont {H.~F.}\ \bibnamefont
			{Hofmann}}, \bibinfo {author} {\bibfnamefont {T.}~\bibnamefont {Nagata}},
		\bibinfo {author} {\bibfnamefont {J.~L.}\ \bibnamefont {O'Brien}}, \bibinfo
		{author} {\bibfnamefont {K.}~\bibnamefont {Sasaki}}, \ and\ \bibinfo {author}
		{\bibfnamefont {S.}~\bibnamefont {Takeuchi}},\ }\href
	{http://iopscience.iop.org/article/10.1088/1367-2630/10/7/073033/meta}
	{\bibfield  {journal} {\bibinfo  {journal} {New J. Phys.}\ }\textbf {\bibinfo
			{volume} {10}},\ \bibinfo {pages} {073033} (\bibinfo {year}
		{2008})}\BibitemShut {NoStop}%
	\bibitem [{\citenamefont {Datta}\ \emph {et~al.}(2011)\citenamefont {Datta},
		\citenamefont {Zhang}, \citenamefont {Thomas-Peter}, \citenamefont {Dorner},
		\citenamefont {Smith},\ and\ \citenamefont {Walmsley}}]{datta11}%
	\BibitemOpen
	\bibfield  {author} {\bibinfo {author} {\bibfnamefont {A.}~\bibnamefont
			{Datta}}, \bibinfo {author} {\bibfnamefont {L.}~\bibnamefont {Zhang}},
		\bibinfo {author} {\bibfnamefont {N.}~\bibnamefont {Thomas-Peter}}, \bibinfo
		{author} {\bibfnamefont {U.}~\bibnamefont {Dorner}}, \bibinfo {author}
		{\bibfnamefont {B.~J.}\ \bibnamefont {Smith}}, \ and\ \bibinfo {author}
		{\bibfnamefont {I.~A.}\ \bibnamefont {Walmsley}},\ }\href
	{http://dx.doi.org/10.1103/PhysRevA.83.063836} {\bibfield  {journal}
		{\bibinfo  {journal} {Phys. Rev. A}\ }\textbf {\bibinfo {volume} {83}},\
		\bibinfo {pages} {063836} (\bibinfo {year} {2011})}\BibitemShut {NoStop}%
	\bibitem [{\citenamefont {Weston}\ \emph {et~al.}(2016)\citenamefont {Weston},
		\citenamefont {Chrzanowski}, \citenamefont {Wollmann}, \citenamefont
		{Boston}, \citenamefont {Ho}, \citenamefont {Shalm}, \citenamefont {Verma},
		\citenamefont {Allman}, \citenamefont {Nam}, \citenamefont {Patel},
		\citenamefont {Slussarenko},\ and\ \citenamefont {Pryde}}]{weston16}%
	\BibitemOpen
	\bibfield  {author} {\bibinfo {author} {\bibfnamefont {M.~M.}\ \bibnamefont
			{Weston}}, \bibinfo {author} {\bibfnamefont {H.~M.}\ \bibnamefont
			{Chrzanowski}}, \bibinfo {author} {\bibfnamefont {S.}~\bibnamefont
			{Wollmann}}, \bibinfo {author} {\bibfnamefont {A.}~\bibnamefont {Boston}},
		\bibinfo {author} {\bibfnamefont {J.}~\bibnamefont {Ho}}, \bibinfo {author}
		{\bibfnamefont {L.~K.}\ \bibnamefont {Shalm}}, \bibinfo {author}
		{\bibfnamefont {V.~B.}\ \bibnamefont {Verma}}, \bibinfo {author}
		{\bibfnamefont {M.~S.}\ \bibnamefont {Allman}}, \bibinfo {author}
		{\bibfnamefont {S.~W.}\ \bibnamefont {Nam}}, \bibinfo {author} {\bibfnamefont
			{R.~B.}\ \bibnamefont {Patel}}, \bibinfo {author} {\bibfnamefont
			{S.}~\bibnamefont {Slussarenko}}, \ and\ \bibinfo {author} {\bibfnamefont
			{G.~J.}\ \bibnamefont {Pryde}},\ }\href {\doibase 10.1364/OE.24.010869}
	{\bibfield  {journal} {\bibinfo  {journal} {Opt. Express}\ }\textbf {\bibinfo
			{volume} {24}},\ \bibinfo {pages} {10869} (\bibinfo {year}
		{2016})}\BibitemShut {NoStop}%
	\bibitem [{\citenamefont {Marsili}\ \emph {et~al.}(2013)\citenamefont
		{Marsili}, \citenamefont {Verma}, \citenamefont {Stern}, \citenamefont
		{Harrington}, \citenamefont {Lita}, \citenamefont {Gerrits}, \citenamefont
		{Vayshenker}, \citenamefont {Baek}, \citenamefont {Shaw}, \citenamefont
		{Mirin},\ and\ \citenamefont {Nam}}]{marsili13}%
	\BibitemOpen
	\bibfield  {author} {\bibinfo {author} {\bibfnamefont {F.}~\bibnamefont
			{Marsili}}, \bibinfo {author} {\bibfnamefont {V.~B.}\ \bibnamefont {Verma}},
		\bibinfo {author} {\bibfnamefont {J.~A.}\ \bibnamefont {Stern}}, \bibinfo
		{author} {\bibfnamefont {S.}~\bibnamefont {Harrington}}, \bibinfo {author}
		{\bibfnamefont {A.~E.}\ \bibnamefont {Lita}}, \bibinfo {author}
		{\bibfnamefont {T.}~\bibnamefont {Gerrits}}, \bibinfo {author} {\bibfnamefont
			{I.}~\bibnamefont {Vayshenker}}, \bibinfo {author} {\bibfnamefont
			{B.}~\bibnamefont {Baek}}, \bibinfo {author} {\bibfnamefont {M.~D.}\
			\bibnamefont {Shaw}}, \bibinfo {author} {\bibfnamefont {R.~P.}\ \bibnamefont
			{Mirin}}, \ and\ \bibinfo {author} {\bibfnamefont {S.~W.}\ \bibnamefont
			{Nam}},\ }\href {http://dx.doi.org/10.1038/nphoton.2013.13} {\bibfield
		{journal} {\bibinfo  {journal} {Nat. Photon.}\ }\textbf {\bibinfo {volume}
			{7}},\ \bibinfo {pages} {210} (\bibinfo {year} {2013})}\BibitemShut {NoStop}%
	\bibitem [{\citenamefont {Klyshko}(1980)}]{klyshko80}%
	\BibitemOpen
	\bibfield  {author} {\bibinfo {author} {\bibfnamefont {D.~N.}\ \bibnamefont
			{Klyshko}},\ }\href {http://stacks.iop.org/0049-1748/10/i=9/a=A09} {\bibfield
		{journal} {\bibinfo  {journal} {Sov. J. Quantum Electron.}\ }\textbf
		{\bibinfo {volume} {10}},\ \bibinfo {pages} {1112} (\bibinfo {year}
		{1980})}\BibitemShut {NoStop}%
	\bibitem [{\citenamefont {Lita}\ \emph {et~al.}(2008)\citenamefont {Lita},
		\citenamefont {Miller},\ and\ \citenamefont {Nam}}]{lita08}%
	\BibitemOpen
	\bibfield  {author} {\bibinfo {author} {\bibfnamefont {A.~E.}\ \bibnamefont
			{Lita}}, \bibinfo {author} {\bibfnamefont {A.~J.}\ \bibnamefont {Miller}}, \
		and\ \bibinfo {author} {\bibfnamefont {S.~W.}\ \bibnamefont {Nam}},\ }\href
	{\doibase 10.1364/OE.16.003032} {\bibfield  {journal} {\bibinfo  {journal}
			{Opt. Express}\ }\textbf {\bibinfo {volume} {16}},\ \bibinfo {pages} {3032}
		(\bibinfo {year} {2008})}\BibitemShut {NoStop}%
	\bibitem [{\citenamefont {Davison}\ and\ \citenamefont
		{Hinkley}(1997)}]{book_davison_1997}%
	\BibitemOpen
	\bibfield  {author} {\bibinfo {author} {\bibfnamefont {A.~C.}\ \bibnamefont
			{Davison}}\ and\ \bibinfo {author} {\bibfnamefont {D.~V.}\ \bibnamefont
			{Hinkley}},\ }\href@noop {} {\emph {\bibinfo {title} {Bootstrap methods and
				their application}}},\ Vol.~\bibinfo {volume} {1}\ (\bibinfo  {publisher}
	{Cambridge Univ. Press},\ \bibinfo {year} {1997})\BibitemShut {NoStop}%
	\bibitem [{\citenamefont {Higgins}\ \emph {et~al.}(2007)\citenamefont
		{Higgins}, \citenamefont {Berry}, \citenamefont {Bartlett}, \citenamefont
		{Wiseman},\ and\ \citenamefont {Pryde}}]{higgins07}%
	\BibitemOpen
	\bibfield  {author} {\bibinfo {author} {\bibfnamefont {B.~L.}\ \bibnamefont
			{Higgins}}, \bibinfo {author} {\bibfnamefont {D.~W.}\ \bibnamefont {Berry}},
		\bibinfo {author} {\bibfnamefont {S.~D.}\ \bibnamefont {Bartlett}}, \bibinfo
		{author} {\bibfnamefont {H.~M.}\ \bibnamefont {Wiseman}}, \ and\ \bibinfo
		{author} {\bibfnamefont {G.~J.}\ \bibnamefont {Pryde}},\ }\href
	{http://dx.doi.org/10.1038/nature06257} {\bibfield  {journal} {\bibinfo
			{journal} {Nature}\ }\textbf {\bibinfo {volume} {450}},\ \bibinfo {pages}
		{393} (\bibinfo {year} {2007})}\BibitemShut {NoStop}%
	\bibitem [{\citenamefont {Matthews}\ \emph {et~al.}(2016)\citenamefont
		{Matthews}, \citenamefont {Zhou}, \citenamefont {Cable}, \citenamefont
		{Shadbolt}, \citenamefont {Saunders}, \citenamefont {Durkin}, \citenamefont
		{Pryde},\ and\ \citenamefont {OТBrien}}]{matthews16}%
	\BibitemOpen
	\bibfield  {author} {\bibinfo {author} {\bibfnamefont {J.~C.~F.}\
			\bibnamefont {Matthews}}, \bibinfo {author} {\bibfnamefont {X.-Q.}\
			\bibnamefont {Zhou}}, \bibinfo {author} {\bibfnamefont {H.}~\bibnamefont
			{Cable}}, \bibinfo {author} {\bibfnamefont {P.~J.}\ \bibnamefont {Shadbolt}},
		\bibinfo {author} {\bibfnamefont {D.~J.}\ \bibnamefont {Saunders}}, \bibinfo
		{author} {\bibfnamefont {G.~A.}\ \bibnamefont {Durkin}}, \bibinfo {author}
		{\bibfnamefont {G.~J.}\ \bibnamefont {Pryde}}, \ and\ \bibinfo {author}
		{\bibfnamefont {J.~L.}\ \bibnamefont {OТBrien}},\ }\href
	{http://dx.doi.org/10.1038/npjqi.2016.23} {\bibfield  {journal} {\bibinfo
			{journal} {{NPJ} {Q}uantum {I}nf.}\ }\textbf {\bibinfo {volume} {2}},\
		\bibinfo {pages} {16023} (\bibinfo {year} {2016})}\BibitemShut {NoStop}%
	\bibitem [{\citenamefont {Harder}\ \emph {et~al.}(2016)\citenamefont {Harder},
		\citenamefont {Bartley}, \citenamefont {Lita}, \citenamefont {Nam},
		\citenamefont {Gerrits},\ and\ \citenamefont {Silberhorn}}]{harder16}%
	\BibitemOpen
	\bibfield  {author} {\bibinfo {author} {\bibfnamefont {G.}~\bibnamefont
			{Harder}}, \bibinfo {author} {\bibfnamefont {T.~J.}\ \bibnamefont {Bartley}},
		\bibinfo {author} {\bibfnamefont {A.~E.}\ \bibnamefont {Lita}}, \bibinfo
		{author} {\bibfnamefont {S.~W.}\ \bibnamefont {Nam}}, \bibinfo {author}
		{\bibfnamefont {T.}~\bibnamefont {Gerrits}}, \ and\ \bibinfo {author}
		{\bibfnamefont {C.}~\bibnamefont {Silberhorn}},\ }\href {\doibase
		10.1103/PhysRevLett.116.143601} {\bibfield  {journal} {\bibinfo  {journal}
			{Phys. Rev. Lett.}\ }\textbf {\bibinfo {volume} {116}},\ \bibinfo {pages}
		{143601} (\bibinfo {year} {2016})}\BibitemShut {NoStop}%
\end{thebibliography}
%

\noindent\textbf{Acknowledgements}
This work is supported by Australian Research Council grant DP140100648. The authors thank Joseph Ho for help with SNSPDs. 
\vspace{1 EM}

\noindent\textbf{Author contributions}

\noindent  G.J.P. conceived the idea and supervised the project. S.S. and M.M.W. constructed and carried out the experiment with help from H.M.C. L.K.S., V.B.V. and S.W.N.  developed the high-efficiency SNSPDs. All authors discussed the results and contributed to the manuscript.

\vspace{1 EM}
\noindent\textbf{Additional information}

\noindent  The data that support the plots within this paper and other findings of this study are available from the corresponding author upon reasonable request. Correspondence and requests for materials should be addressed to G.J.P.

\vspace{1 EM}

\noindent\textbf{Competing financial interests}

\noindent The authors declare no competing financial interests.

\end{document}